\newcommand{\CLASSINPUTbottomtextmargin}{0.95in}
\documentclass[conference]{IEEEtran}
\IEEEoverridecommandlockouts
\usepackage{amsfonts}
\usepackage{dsfont} 
\usepackage{setspace}
\usepackage{color}
\usepackage{amssymb}
\usepackage{cite}
\usepackage[cmex10]{amsmath}
\usepackage{algorithm}
\usepackage{algorithmic} % algorithm 
\usepackage{array}
\usepackage{mathrsfs}
\usepackage{graphicx}
\usepackage{latexsym}
\usepackage{amscd}
\usepackage{amsfonts}
\usepackage{subfigure}
\usepackage{amsmath,amscd,amssymb,verbatim}
\usepackage{graphics}
\usepackage{amsthm}
\usepackage[T1]{fontenc}
\usepackage[utf8]{inputenc}
\usepackage{authblk}
\usepackage[nocomma]{optidef}
\usepackage{mathtools}
\usepackage{anyfontsize} 
\usepackage{soul}
\usepackage{textcomp}
\usepackage{xcolor}
\usepackage{multirow}
\usepackage{amsmath}
\usepackage{blindtext}
\usepackage{enumitem}

\IEEEoverridecommandlockouts

\newtheorem{lemma}{\underline{Lemma}}%[section]

\begin{document}
\setlength{\columnsep}{0.2in}
\title{ 
{
Efficient Joint Resource Allocation for Wireless Powered ISAC with Target Localization} 
\vspace{-0.2cm}}

\author[$\dag$]{Boyao Li$^{\dag}$, Qinwei He$^\ddag$, Boao Zhang$^{\dag}$, Xiaopeng Yuan$^{\dag*}$ and Anke Schmeink$^{\dag}$\\
$^\dag$INDA Chair, RWTH Aachen University, Germany, Email: $li|yuan|schmeink$@inda.rwth-aachen.de\\
$^\ddag$Global Energy Interconnection Research Institute Europe GmbH, Germany, Email: $qinwei.he$@geiri.eu
\vspace{-0.3cm}
\thanks{%The work of  Y. Hu and B. Ai is supported by the National Key R\&D Program of China under Grant 2021YFB2900301, 2023YFE0206600, and the  Fundamental Research Funds for the Central Universities (2042024kf1006). The work of B. Li, X. Yuan and A. Schmeink is supported by BMBF Germany in the program of ``Souveran. Digital. Vernetzt.'' Joint Project 6G-RIC with number 16KISK028.
The work %of B. Li, B. Zhang, X. Yuan, and A. Schmeink 
was supported in part by BMFTR Germany in the projects 6G-RIC under Grant 16KISK028, 6GEM+ under Grant 16KIS2409K, and GEM-X under Grant 16KISS004K, and in part by the State Grid Corporation of China under Grant 5700-202358798A-3-9-HW.} %The work of Q. He was supported by the State Grid Corporation of China under Grant 5700-202358798A-3-9-HW. 
\thanks{$^*$X. Yuan is the corresponding author.}
}
  %\emph{Member, IEEE}
\vspace{-1cm}
\IEEEoverridecommandlockouts
    \IEEEpubid{\begin{minipage}{\textwidth}\ \\[50pt]
            {\copyright 2026 IEEE. Personal use of this material is permitted.  Permission from IEEE must be obtained for all other uses, in any current or future media, including reprinting/republishing this material for advertising or promotional purposes, creating new collective works, for resale or redistribution to servers or lists, or reuse of any copyrighted component of this work in other works. %Citation information: DOI 10.1109/TWC.2025.3635227
            }
    \end{minipage}} 
\maketitle
%\IEEEpubidadjcol

%%%%%%%%%%%%%%%%%%%%%%%%%%%%%%%%%%%%%%%%%%%%%%%%%%%%%%%%%%%%%
\begin{abstract}
Integrated sensing and communication (ISAC) enables 6G networks to share radio resources for sensing and data transmission. For energy-limited ISAC systems, wireless power transfer (WPT) can be integrated as a power supply, which also introduces tradeoffs among harvested energy, communication throughput, and sensing accuracy. Motivated by this, we study a wireless powered multi-user ISAC system for multi-target localization, where a base station (BS) broadcasts a WPT signal for energy harvesting and users then transmit data with the harvested energy to the BS sequentially. Meanwhile, both the WPT signal and the users' data signals are exploited for target localization at the BS. By imposing Cram\'er-Rao bound (CRB) constraints to guarantee localization accuracy, we formulate a minimum user throughput maximization problem that jointly optimizes the WPT duration, per-user transmission durations, and user transmit powers. To address the variable coupling and inherent nonconvexity of the problem, we derive a more tractable reformulation via variable substitutions and logarithmic objective transformations. Since the CRB constraints remain nonconvex, we further develop a successive convex approximation (SCA)-based iterative algorithm and 
finally obtain an efficient suboptimal solution. Numerical results demonstrate the algorithm convergence and significant performance gains of the proposed scheme over benchmark schemes, highlighting the importance of joint resource allocation design for wireless powered ISAC systems.
\end{abstract}

%\vspace{-.2cm}
% \vspace{-.4cm}
\begin{IEEEkeywords}
%\vspace{-.2cm}
Integrated sensing and communication (ISAC), wireless power transfer (WPT), target localization, Cram\'er-Rao bound (CRB), efficient resource allocation.
\end{IEEEkeywords}

%\vspace{-.2cm}
%\newpage
%%%%%%%%%%%%%%%%%%%%%%%%%%%%%%%%%%%%%%%%%%%%%%%%%%%%%%%%%%%%%
%\vspace{-0.2cm}
\section{Introduction}
Integrated sensing and communication (ISAC) has emerged as a key pillar of 6G wireless networks, enabling high-quality wireless connectivity and accurate sensing capability within a unified infrastructure \cite{ISAC_advance}. By sharing spectrum and hardware%, %and waveforms
, ISAC is 
expected to 
support %rapidly %a variety of 
%emerging 
diverse applications %such as 
%including 
such as smart home, human-computer interaction, and vehicle-to-everything (V2X)
\cite{ISAC_application}. This tight integration, however, also introduces %fundamental 
tradeoffs between sensing and communication due to %, since %both functionalities must %share 
shared limited resources.% are used to satisfy highly heterogeneous performance requirements. 

To %address 
navigate these tradeoffs, extensive research efforts have been devoted to ISAC%system designs
, e.g., in waveform design \cite{ISAC_waveform1,ISAC_waveform2} and advanced multiple access schemes \cite{ISAC_multiple_access}. %multi-input multi-output (MIMO) techniques
%signal processing \cite{ISAC_signal}.
%, as well as performance analysis based on sensing-oriented metrics. 
In accordance with different sensing 
demands, various sensing tasks have been studied, including detection, %parameter
estimation, and %activity
recognition \cite{ISAC_dual}. In particular, target localization, as a representative estimation task, is crucial for %enabling 
location-aware services such as %like
indoor navigation and tracking.
%Among these tasks, target localization is particularly important as it directly enables location-aware services.
This has %motivated 
spurred increasing interest in 
%integrating target localization into communication
localization-oriented ISAC, e.g., in cooperative networks \cite{ISAC_localization1} and unmanned aerial vehicle (UAV)-assisted scenarios \cite{ISAC_localization2}. 
%localization-oriented ISAC designs, where the same radio infrastructure and signals are jointly leveraged for data transmission and positioning.
%Among them, target localization has attracted increasing attention due to its essential role in many location-aware services, and correspondingly, a performance metric Cram\'er-Rao bound (CRB) has been widely adopted.
%As one of the two core functionalities of ISAC, sensing can be broadly categorized into three task types: detection, estimation, and recognition \cite{ISAC_dual}. Detection-oriented sensing determines whether a target is present, whereas recognition focuses on identifying the target category or event type. In contrast, estimation aims to extract parameters of interest from the sensed object, such as angle, velocity, or location, and has become increasingly attractive in both academia and industry. Estimation performance is commonly quantified by the mean squared error (MSE). %, defined as the expected squared difference between the true and estimated parameters \cite{estimation_book}. 
% When the exact MSE is difficult to obtain, an informative alternative is the Cram\'er-Rao bound (CRB), which provides a lower bound on the error variance achievable by unbiased estimators \cite{estimation_book,MSE_CRB}. In particular, for target localization, the CRB is widely adopted to characterize the attainable positioning accuracy \cite{target_localization}, since it explicitly captures the impact of sensing geometry.

Despite its potential, %practical 
ISAC deployment becomes challenging %are often 
in energy-limited scenarios. %constrained by  energy supply
%, as many ISAC devices, e.g., sensors and edge nodes, are powered by finite-capacity batteries.
To address this challenge, wireless power transfer (WPT), %can eliminate the need for frequent battery replacement and 
%significantly prolong the operational lifetime of energy-limited wireless devices, 
by harvesting energy from ambient or dedicated radio frequency (RF) signals, %. Therefore, WPT 
has %therefore 
%been introduced as a promising solution 
become a natural solution to enable sustainable ISAC operation %by replenishing energy at 
for %energy-limited 
batteryless devices % in remote areas 
\cite{ISCPT,ISCAP}. %In a typical WPT-enabled system, a dedicated transmitter (e.g., the BS) first broadcasts an energy-carrying waveform in the downlink, and energy-limited devices harvest the received radio-frequency (RF) energy to power their subsequent transmissions. 
%Under a %finite total time budget
%latency requirement
%In latency-critical applications,
%Under stringent time
Under a finite time budget, extending the WPT phase increases the harvested energy, but inevitably reduces the  available time for the subsequent ISAC phase.
To %fully exploit the harvested energy
optimize this tradeoff, efficient resource allocation is crucial to %balance %sensing accuracy 
%target localization 
%and communication throughput
%time scheduling 
orchestrate the energy-communication-localization interplay in wireless powered ISAC systems.

To this end, several works have investigated resource allocation for wireless powered ISAC. In \cite{wireless_ISAC}, the authors jointly optimize power control as well as WPT and ISAC beamforming to improve sensing performance while guaranteeing the communication %performance constraints
requirements. %Along a different line
In addition, \cite{UAV_ISAC} investigates a UAV-enabled wireless powered ISAC system and jointly optimizes the filter and waveform design, time scheduling and uplink powers of users, and UAV trajectory. % via a multi-objective design.%studies a UAV-enabled wireless powered ISAC architecture with time scheduling and joint resource optimization. %by designing the sensing/WPT waveform together with uplink transmission parameters .%studies UAV-enabled wireless powered ISAC with %TDMA-based 
%time scheduling and joint resource optimization. %, while characterizing sensing performance via a signal-to-interference-plus-noise ratio (SINR) metric, which is insufficient for target localization. 
%However, the WPT and ISAC durations are fixed, and the WPT phase is used purely for energy delivery without contributing to sensing, leaving the potential gains from time optimization and WPT-sensing co-design unexplored.
However, these designs do not explicitly consider the target
localization as the sensing task. Their sensing objectives are %mainly
characterized by %radar-oriented
nonspecific metrics such as %mean square error (MSE) or 
signal-to-interference-plus-noise ratio (SINR), 
rather than localization-specific metrics such as Cram\'er-Rao bound (CRB) that directly quantifies %positioning
localization accuracy. Moreover, they do not fully exploit the sensing opportunities from all phases and nodes, and thus cannot leverage the sensing geometry required for accurate %multi-target 
localization. %(e.g., jointly leveraging both WPT and ISAC signals and the resulting multi-static paths)
% , thereby leaving the %
% potential gains from joint resource allocation of wireless powered ISAC with target localization with full system exploitment
% %localization resource allocation and WPT-sensing co-design for multi-target localization %insufficiently 
% %the study of 
% unexplored.
Consequently, the potential gains from %localization-guaranteed
WPT-ISAC co-design and CRB-constrained resource allocation remain insufficiently explored.

To fill this gap, we study a wireless powered multi-user ISAC system, % with multiple sensing targets, %where the WPT signal is exploited for both power transfer and sensing
where the base station (BS) first performs WPT to energize the users, and then users transmit data to the BS with the obtained energy, while both phases jointly contribute to the multi-target localization task guaranteed via CRB constraints. 
The main contributions of this paper are summarized as follows:
\begin{itemize}
    \item \textbf{Localization-specific wireless powered ISAC framework:} We propose a two-phase WPT-ISAC protocol %system 
    for multi-target localization in energy-limited systems, where the localization geometry and opportunities are fully considered and exploited. %, where the BS first broadcasts a downlink WPT signal to charge users, and the users subsequently utilize the harvested energy to transmit uplink data to the BS in a TDMA manner. Meanwhile, 
    %Both the WPT signal and the users' uplink data signals are exploited to localize the targets at the BS. 
    Under this %scheme
    design, we formulate a max-min throughput problem by jointly optimizing the WPT duration, per-user transmission time and power, subject to energy causality and per-target CRB-based localization accuracy constraints. The formulation %which
    explicitly captures the %coupled 
    tradeoffs among harvested energy, communication performance, and localization accuracy.
%We develop a wireless powered multi-user ISAC framework for multi-target localization, where the BS performs downlink WPT to sustain energy for users's uplink communication. Both the WPT  and ISAC waveforms are exploited for target localization. This model captures a practically important tradeoffs in between energy delivery, uplink communication, and target localization.%We formulate a joint optimization problem to maximize the minimum user throughput by optimizing the WPT/ISAC time allocation and transmit power, under energy causality and per-target CRB-based localization accuracy constraints.%We propose a wireless powered ISAC framework specialized for target localization, where the WPT phase contributes to sensing in addition to energy transfer, and the localization accuracy is guaranteed via CRB constraints.
    \item \textbf{%Problem Reformulation and 
    %CRB-constrained 
    Efficient 
    joint resource allocation:}  To tackle the resulting nonconvex time-power coupled problem with intricate CRB constraints, we develop an equivalent reformulation via variable substitutions and %a 
    logarithmic transformations, which %decouples the %coupling
    %multiplication structure
    removes the multiplicative coupling. %Building on 
    After the reformulation, we construct convex surrogates for the remaining nonconvexity in CRB constraints and propose an efficient successive convex approximation (SCA)-based iterative algorithm to obtain a high-quality suboptimal solution.%To make the resulting nonconvex problem more tractable, we first introduce variable transformations and exploit the monotonicity of logarithmic functions to simplify the original coupling structure. Building on the reformulation, we address the remaining nonconvexity through convex approximation, resulting in an effective iterative algorithm that yields a high-quality suboptimal solution.
    \item %\textbf{Numerical Investigation:} Numerical results verify that the proposed joint optimization scheme significantly outperforms benchmark schemes, highlighting the importance of coordinated time and power allocation in wireless powered ISAC systems.
    \textbf{Numerical investigation%in wireless powered ISAC
    :}
    Numerical results demonstrate fast convergence of the iterative algorithm and significant throughput gains of the proposed scheme over benchmark schemes, and provide insights into how resources can be jointly allocated to balance heterogeneous performance requirements in wireless powered ISAC systems.
\end{itemize}
The rest of this paper is organized as follows. In Section II, we introduce the system model and the CRB model, and formulate the optimization problem. Section III develops the proposed iterative algorithm. Numerical results are presented in Section IV, and conclusions are drawn in Section V.

\section{Problem Formulation}
\subsection{System Model}
\vspace{-0.1cm}
\begin{figure}[!t]
	\centering
	\includegraphics[width=0.48\textwidth, trim=10 50 10 20]{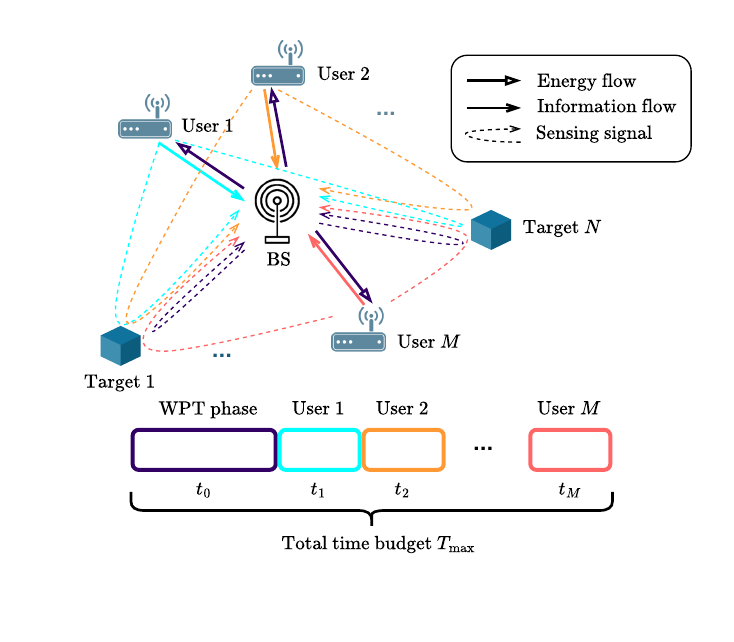}
	\caption{Illustration of a wireless powered $M$-user ISAC system with $N$ targets under a total time budget $T_{\mathrm{max}}$.}
    \label{ISAC_scenario}
	\vspace{-0.6cm}
\end{figure}
We consider a wireless powered ISAC system consisting of $M$ batteryless users, $N$ passive targets, and a BS located at $\mathbf{x}_0 \in \mathbb{R}^2$, as shown in Fig.~\ref{ISAC_scenario}. The position of user $m \in \mathcal{M} \triangleq \{1,\ldots,M\}$ is denoted by $\mathbf{x}_m \in \mathbb{R}^2$, while the reference position %\footnote{Here, the position refers to a reference target point corresponding to a preliminary location estimate, which is assumed to be available from prior sensing results.} 
of target $n \in \mathcal{N} \triangleq \{1,\ldots,N\}$ is denoted by $\mathbf{q}_n \in \mathbb{R}^2$. The users aim to transmit %uplink 
data to the BS, while the BS is required to localize the passive targets based on the received sensing signals.%, %which is assumed to be available 
%obtained from prior sensing results.

In the WPT phase, the BS broadcasts an energy-carrying %waveform
signal with %fixed 
power $p_0$ over a duration $t_0$, from which all users harvest energy for the subsequent ISAC phase. 
The harvested energy at user $m$ is given by
\begin{equation}
\label{Em}
    E_m(t_0) = \zeta_m h_{m} t_0 p_0,~\forall m\in\mathcal{M},
\end{equation}
where $\zeta_m$ is the energy conversion efficiency of user $m$ and $h_m$ denotes the channel gain from the BS to user $m$.

After the WPT phase, the system enters an ISAC phase consisting of $M$ %time division multiple access (TDMA) 
slots. In the $m$-th slot, user $m$ transmits an information-bearing waveform with power $p_m$ over a duration $t_m$, which simultaneously supports target localization and uplink communication. 

During the ISAC phase, while illuminating the targets for sensing,
user $m$ simultaneously transmits its data to the BS. In slot $m$, the signal-to-noise ratio (SNR) at the BS is
\begin{equation}
    \gamma_m(p_m) = \frac{p_m h_m}{\sigma^2},~\forall m\in\mathcal{M}.
\end{equation}
%Under the Shannon capacity model, 
The achievable throughput of user $m$ for a transmission duration $t_m$ is expressed as
\begin{equation}
    R_m(t_m,p_m) = t_m W\log_2(1 + \gamma_m(p_m)),~\forall m\in\mathcal{M}.\label{Rm}
\end{equation}

The total transmission duration of the WPT phase and all ISAC slots is constrained by a total time budget $T_{\mathrm{max}}$, i.e.,
\begin{equation}
    \sum_{m=0}^{M} t_m \le T_{\mathrm{max}}.
    \label{con_t}
\end{equation}
Since the transmission of batteryless user $m$ is solely powered by the harvested energy $E_m(t_0)$, the following energy causality constraint must hold:
\begin{equation}
     t_m p_m\le E_m(t_0),~\forall m\in\mathcal{M}.
     \label{con_E}
\end{equation}
Furthermore, due to hardware limitations, the transmit power of each user is bounded by
\begin{equation}
    0 < p_m \le P_{\mathrm{max}},~\forall m\in\mathcal{M}.
    \label{con_p}
\end{equation}

\subsection{CRB-Based Localization Model}
\vspace{-0.1cm}
To quantify the target localization accuracy, we adopt a CRB-based localization model. Both the WPT signal and the users' ISAC signals illuminate all targets, which then backscatter the incident energy toward the BS. The BS receives the echo signals and processes them to estimate the target positions. Thus, each user provides a distinct bistatic sensing geometry, while the BS provides a %complementary 
monostatic geometry for localization.%Thus, each user provides an independent sensing view of the targets.
%This slot-by-slot illumination enables each user to contribute to the sensing process.

For each target $n\in\mathcal{N}$, we characterize both the bistatic and monostatic sensing geometries. The bistatic round-trip distance in slot $m$ is given by
\begin{equation}
r_{m,n} = \|\mathbf{x}_m - \mathbf{q}_n\| + \|\mathbf{x}_0 - \mathbf{q}_n\|,~\forall m\in\mathcal{M},~\forall n\in\mathcal{N},
\end{equation}
while the monostatic round-trip distance is
\begin{equation}
r_{0,n} = 2\|\mathbf{x}_0 - \mathbf{q}_n\|,~\forall n\in\mathcal{N}.
\end{equation}
%For each $n\in\mathcal{N}$, 
Taking the gradient of $r_{m,n}$ with respect to the reference target location $\mathbf{q}_n$
yields
\begin{align}
\label{eq_XY}
\nabla_{\mathbf{q}_n} r_{m,n}
&=
\begin{cases}
-\dfrac{\mathbf{x}_m - \mathbf{q}_n}{\|\mathbf{x}_m - \mathbf{q}_n\|}
-\dfrac{\mathbf{x}_0 - \mathbf{q}_n}{\|\mathbf{x}_0 - \mathbf{q}_n\|},
& m \in \mathcal{M}, \\[10pt]
-2\dfrac{\mathbf{x}_0 - \mathbf{q}_n}{\|\mathbf{x}_0 - \mathbf{q}_n\|},
& m = 0,
\end{cases}\nonumber\\
&\triangleq
\begin{bmatrix}
X_{m,n} \\ 
Y_{m,n}
\end{bmatrix},~\forall m\in\mathcal{M}\cup\{0\},~\forall n\in\mathcal{N}.
\end{align}
Following the classical CRB analysis for range-based target localization \cite{target_localization,target_localization_con}, the obtained gradient terms in \eqref{eq_XY} are used to construct the Fisher information matrix (FIM) associated with target $n\in\mathcal{N}$, which can be
expressed as
\begin{equation}
\label{FIM}
\mathbf{J}_n= 
\sum_{m=0}^{M} p_m K_{m,n}
\begin{bmatrix}
X_{m,n}^{2} & X_{m,n}Y_{m,n} \\
X_{m,n}Y_{m,n} & Y_{m,n}^{2}
\end{bmatrix},%~\forall n\in\mathcal{N},
\end{equation}
where the scaling factor $K_{m,n}$ is defined as
\begin{equation}
K_{m,n}
=
\frac{8\pi^{2}W^{2}h_{m,n}}{\sigma^{2}c^{2}},~\forall m\in\mathcal{M}\cup\{0\},~\forall n\in\mathcal{N}.
\end{equation}
Here, $W$ denotes the effective bandwidth, $h_{m,n}$ denotes the channel gain from transmitter $m$ to target $n$, with
$m=0$ corresponding to the BS and $m \in \mathcal{M}$ corresponding to the users. Moreover, $\sigma^2$
is the noise power at the BS and $c$ is the speed of light.

For notational convenience, the FIM associated with target $n$ in \eqref{FIM} is rewritten in the compact form:
\begin{equation}
\mathbf{J}_n
=
\begin{bmatrix}
A_n & C_n \\
C_n & B_n
\end{bmatrix},~\forall n\in\mathcal{N},
\end{equation}
where
\begin{align}
    A_n &= \sum_{m=0}^{M} p_m K_{m,n} X_{m,n}^{2},\quad B_n = \sum_{m=0}^{M} p_m K_{m,n} Y_{m,n}^{2},\nonumber\\
    C_n &= \sum_{m=0}^{M} p_m K_{m,n} X_{m,n}Y_{m,n}.
\end{align}
Under non-degenerate %sensing
localization geometry, the FIM is positive definite, i.e., $A_n B_n - C_n^2 > 0$ %. Therefore, 
and the CRB matrix can be expressed as the inverse of FIM: %as%$\mathbf{J}_n$:
\begin{equation}
    \mathbf{J}_n^{-1}
    =
    \frac{1}{A_n B_n - C_n^{2}}
    \begin{bmatrix}
    B_n & -C_n \\
    -C_n & A_n
    \end{bmatrix},~\forall n\in\mathcal{N}.
\end{equation}
The trace of the CRB matrix represents a lower bound %on the sum of the MSEs 
for the target location estimation and is given by
\begin{equation}
    \mathrm{tr}(\mathbf{J}_n^{-1})= \frac{A_n + B_n}{A_n B_n - C_n^{2}},~\forall n\in\mathcal{N}.
\end{equation}
Given a required localization accuracy threshold $\eta>0$, the following constraint must be satisfied:
\begin{equation}
    \mathrm{tr}(\mathbf{J}_n^{-1}) \le \eta,~\forall n\in\mathcal{N}.
    \label{con_CRB}
\end{equation}

To maximize the communication performance while ensuring fairness among all users, the minimum achievable throughput is adopted as the optimization objective. Accordingly, the overall optimization problem is formulated as
\vspace{-0.1cm}
\begin{subequations}
    \begin{align}(\mathcal{P}1)\!:\max _{\mathbf{t},\mathbf{p}} ~~&\min_{m\in\mathcal{M}}\{R_m(t_m,p_m)\}\nonumber\label{P1_obj}\\
    \text{ s.t. }~~
    &t_m >0,~\forall m\in\mathcal{M}\cup\{0\},\\
    % &\sum_{m=0}^Mt_m\leq T_{\mathrm{max}},\\
    % &0 < p_m \leq P_{\mathrm{max}},~\forall m\in\mathcal{M},\\
    % & t_m p_m  \le E_m(t_0),~\forall m\in\mathcal{M},\label{P1_energy_con}\\   
    % & \mathrm{tr}(\mathbf{J}_n^{-1}) \le \eta,~\forall n\in\mathcal{N},%F_n(\mathbf{p}) \le 0,~\forall n\in\mathcal{N}, 
    &\eqref{con_t},\eqref{con_E},\eqref{con_p},\eqref{con_CRB},\nonumber
    \label{P1_ABC}
    \end{align}
\end{subequations}
where $\mathbf{t} \triangleq [t_0,t_1,\ldots,t_M]^{\mathsf T}\in\mathbb{R}^{M+1}$ collects the transmit durations of the BS and all users. Problem $(\mathcal{P}1)$ is nonconvex since
$R_m(t_m,p_m)$ is %neither jointly concave nor jointly convex with respect to
not jointly concave in $(t_m,p_m)$ %, as indicated by its indefinite Hessian matrix. 
and nonconvex constraints %\eqref{P1_energy_con}-\eqref{P1_ABC} 
\eqref{con_E} and \eqref{con_CRB} involve
multiplicative couplings among the optimization variables. As a result, problem $(\mathcal{P}1)$ cannot be directly solved using standard
convex optimization techniques.

\section{Proposed Iterative Solution}
This section develops an efficient iterative algorithm to solve $(\mathcal{P}1)$. 
We first reformulate $(\mathcal{P}1)$ via variable substitutions and logarithmic transformations, and then apply SCA to handle the remaining nonconvexity.

\subsection{Problem Reformulation}
\vspace{-0.1cm}

To handle the multiplicative coupling between $t_m$ and $p_m$ in constraints %\eqref{P1_energy_con} and \eqref{P1_ABC}
\eqref{con_E} and \eqref{con_CRB}, %First, 
%we introduce new optimization variables
we apply a logarithmic change of variables by replacing $(t_m,p_m)$ with $(u_m,v_m)$ as%by replacing ($t_m,p_m$) 
\begin{equation}
u_m = \log(t_m),~v_m = \log(p_m),~\forall m\in\mathcal{M},
\end{equation}
which guarantees $t_m>0$ and $p_m>0$ automatically. Under this substitution, %taking the logarithm on both sides of %
the energy causality constraint
\eqref{con_E} %can be equivalently written as
%yields
becomes
\begin{equation}
\label{u_plus_v}
u_m + v_m \le \log(E_m(t_0)),~\forall m\in\mathcal{M},
\end{equation}
which converts the original product terms $t_mp_m$ into a linear term.
Moreover, since $E_m(t_0)$ %defined 
in \eqref{Em} is affine in $t_0$, $\log(E_m(t_0))$ is concave in $t_0$, and thus constraint \eqref{u_plus_v} is convex. Also, with variable replacement, the throughput \eqref{Rm} transforms to
\begin{equation}
    \widetilde{R}_m(u_m,v_m) = e^{u_m} W\log_2\Big(1 + \frac{h_{m}}{\sigma^2} e^{v_m}\Big),~\forall m\in \mathcal{M},
\end{equation}
%R_m(t_m,p_m) 
%&
%\end{align}
which is still not jointly concave in $(u_m,v_m)$.

To overcome this difficulty, we apply a logarithmic transformation to $\widetilde{R}_m(u_m,v_m)$, inspired by a similar technique to that in our prior works \cite{loglog1,loglog2}. Since the logarithm is strictly increasing, maximizing $\widetilde{R}_m(u_m,v_m)$ is equivalent to 
maximizing $\log \widetilde{R}_m(u_m,v_m)$, which can be expressed as
\begin{align}
&\log \widetilde{R}_m(u_m,v_m)\\
=\;&
u_m + \log(W)+\log\Big(\log_2\Big(1+\frac{h_{m}}{\sigma^2} e^{v_m}\Big)\Big),~\forall m\in\mathcal{M}.\nonumber
\end{align}
The first term is affine in $u_m$ and $\log(W)$ is a constant, while the last term is concave in $v_m$, as established in the following lemma.
%\vspace{-0.2cm}
\begin{lemma}
\label{lem:concavity_vm}
For any $h_{m} > 0$ and $\sigma^2 > 0$, the function
\begin{equation}
f(v_m) = \log\left(\log_2\left(1 + \frac{h_{m}}{\sigma^2} e^{v_m}\right)\right)
\end{equation}
is strictly concave in $v_m \in \mathbb{R}$.
%\vspace{-0.3cm}
\end{lemma}

\begin{proof}
To prove the concavity of $f(v_m)$,
% \begin{equation}
%     f(v_m)=\log\Big(\log_2\Big(1+\frac{h_{0,m}}{\sigma^2} e^{v_m}\Big)\Big),
% \end{equation}
we define $k_m=\frac{h_{m}}{\sigma^2}> 0$ and 
$g(v_m)=1+k_m e^{v_m}>1$.  
Using the identity 
$\log_2 g(v_m)=\frac{\log g(v_m)}{\log 2}$, the function $f(v_m)$ becomes
\begin{equation}
    f(v_m)=\log(\log g(v_m)) - \log(\log 2).
\end{equation}
%Since $\log(\log 2)$ is constant, its derivative vanishes.  
The first derivative of $f(v_m)$ is therefore
\begin{equation}
f'(v_m)
= \frac{k_m e^{v_m}}
{g(v_m)\,\log g(v_m)}.
\end{equation}
Differentiating once more yields
\begin{align}
&f''(v_m)\nonumber\\
=\;&
\frac{
k_m e^{v_m}
\big(g(v_m)\log g(v_m) 
      - k_m e^{v_m}(\log g(v_m)+1)\big)
}
{g(v_m)^2 (\log g(v_m))^2}\nonumber\\
=\;&
\frac{
k_m e^{v_m}(\log g(v_m) 
      - k_m e^{v_m})
}
{g(v_m)^2 (\log g(v_m))^2},
\end{align}
Since $g(v_m)>1$, using the %we have $k_m e^{v_m}=g(v_m)-1$, and the
well-known inequality
\begin{equation}
\log(y)< y-1,~\forall y>1,
\end{equation}
we have
\begin{equation}
\log g(v_m) < g(v_m)-1=k_m e^{v_m}.
\end{equation}
Hence, the numerator of $f''(v_m)$ is negative, while the denominator is positive, and therefore $f''(v_m)< 0$ for all $v_m\in \mathbb{R}$, which completes the proof.
%Thus, $f(v_m)$ is strictly concave.
\end{proof}
Consequently, $\log \widetilde{R}_m(u_m,v_m)$ is jointly concave in $(u_m,v_m)$, as the sum of an affine function in $u_m$ and a concave function in $v_m$. Moreover, the objective function $\min_{m\in\mathcal{M}}\{\log \widetilde{R}_m(u_m,v_m)\}$ is concave as the pointwise minimum of concave functions.%remains concave, as the pointwise minimum of concave functions is also concave.

As for CRB constraint \eqref{con_CRB}, since $A_n B_n - C_n^{2}>0$, the above inequality can be rewritten as
\begin{equation}
\label{CRB_con}
A_n + B_n - \eta(A_n B_n - C_n^2) \le 0,~\forall n\in\mathcal{N}.
\end{equation}

To reveal the algebraic structure of $A_n B_n - C_n^{2}$ and its dependence on the power variables $\{p_m\}$, for each $n\in\mathcal{N}$, we define
$a_{m,n}=\sqrt{p_m K_{m,n}}\,X_{m,n}$, $b_{m,n}=\sqrt{p_m K_{m,n}}\,Y_{m,n}$. %therefore %$\mathbf{a}_n =[a_{0,n},\ldots,a_{M,n}]^{\mathsf T}$ and $\mathbf{b}_n =[b_{0,n},\ldots,b_{M,n}]^{\mathsf T}$.
%then $A_n=\|\mathbf{a}_n\|^{2}$, $B_n=\|\mathbf{b}_n\|^{2}$, and 
%$C_n=\mathbf{a}_n^{\mathsf T}\mathbf{b}_n$.  
%Hence, %for all $n\in\mathcal{N}$,
With these definitions, $A_n=\sum_{m=0}^{M} a_{m,n}^{2}$, $B_n=\sum_{m=0}^{M} b_{m,n}^{2}$, and $C_n=\sum_{m=0}^{M} a_{m,n}b_{m,n}$. Applying Lagrange’s identity, we obtain
\begin{align}
\label{ABC_schwarz}
&A_n B_n - C_n^{2}\nonumber\\
%=\;& \|\mathbf{a}_n\|^{2}\|\mathbf{b}_n\|^{2} - (\mathbf{a}_n^{\mathsf T}\mathbf{b}_n)^{2}\nonumber\\
=\;& \frac12 \sum_{i=0}^{M}\sum_{j=0}^{M} (a_{i,n} b_{j,n} - a_{j,n} b_{i,n})^{2}\nonumber\\
=\;& \frac12 \sum_{i=0}^{M}\sum_{j=0}^{M} p_i p_j K_{i,n}K_{j,n}(X_{i,n}Y_{j,n}-X_{j,n}Y_{i,n})^{2}.
\end{align}
%where the second equality follows from a standard algebraic identity
%related to the Cauchy-Schwarz inequality.
 
By substituting \eqref{ABC_schwarz} and separating the terms involving the fixed BS power $p_0$, the CRB constraint \eqref{CRB_con} can be equivalently reformulated as
\begin{align}
\label{Fn}
F_n(\mathbf{p})=&\sum_{m=1}^{M} \alpha_{m,n} p_m + \mu_n
-\frac{\eta}{2}\sum_{i=1}^{M}\sum_{j=1}^{M}
\beta_{i,j,n} p_i p_j\nonumber\\
&-\eta p_0 \sum_{m=1}^{M} \varphi_{m,n} p_m \leq 0,~
\forall n\in\mathcal{N},
\end{align}
% \begin{align}
% &A_n + B_n - \eta(A_n B_n - C_n^2)\nonumber\\
% =&\sum_{m=1}^{M} \alpha_{m,n} p_m + \mu_n
% -\frac{\eta}{2}\sum_{i=1}^{M}\sum_{j=1}^{M}
% \beta_{i,j,n} p_i p_j\nonumber\\
% &-\eta p_0 \sum_{m=1}^{M} \varphi_{m,n} p_m\nonumber\\
% \triangleq&
% F_n(\mathbf{p}),~
% \forall n\in\mathcal{N},
% \end{align}
where $\mathbf{p} \triangleq [p_1,\ldots,p_M]^{\mathsf T}$ denotes the vector of transmit powers. The coefficients for each $n\in\mathcal{N}$
%$\alpha_{m,n}$, $\mu_n$, $\beta_{i,j,n}$, and $\varphi_{m,n}$ 
are defined as
\begin{align}
\!\!\alpha_{m,n}
&\triangleq
K_{m,n}(X_{m,n}^{2}+Y_{m,n}^{2}),~m\in\mathcal{M},\\
\!\!\mu_n
&\triangleq p_0 K_{0,n}(X_{0,n}^2+Y_{0,n}^2),\\
\!\!\beta_{i,j,n}
&\triangleq
K_{i,n}K_{j,n}
(X_{i,n}Y_{j,n}-X_{j,n}Y_{i,n})^{2},
~i,j\in\mathcal{M},\\
\!\!\varphi_{m,n}
&\triangleq K_{0,n}K_{m,n}(X_{0,n}Y_{m,n}-\!X_{m,n}Y_{0,n})^{2},\,m\in\mathcal{M}.
\end{align}
%By 
Replacing the optimization variables, \eqref{Fn} can be %compactly 
written as
\begin{align}
\label{eq:Fn_equiv}
%F_n(\mathbf{p})
\widetilde{F}_n(\mathbf{v})=&
\sum_{m=1}^{M} \alpha_{m,n} e^{v_m}
+\mu_n
-\frac{\eta}{2}\sum_{i=1}^{M}\sum_{j=1}^{M}
\beta_{i,j,n} e^{v_i + v_j}\nonumber\\
&-\eta p_0 \sum_{m=1}^{M} \varphi_{m,n} e^{v_m}
%\triangleq
,~\forall n\in\mathcal{N}.
\end{align}
While the first term is convex in $\mathbf{v}$ and $\mu_n$ is constant, the last two terms are concave because $e^{v_i+v_j}$ and $e^{v_m}$ are convex in $\mathbf{v}$, and the negative of a convex function is concave. Thus, $\widetilde{F}_n(\mathbf{v})$ is nonconvex.

Overall, the reformulated optimization problem is given by
\begin{subequations}
    \begin{align}(\mathcal{P}2)\!:\max _{t_0,\mathbf{u},\mathbf{v}} ~~&\min_{m\in\mathcal{M}}\{\log \widetilde{R}_m(u_m,v_m)\}\nonumber\\
    \text{ s.t. }~~
    &t_0 >0,\label{P2_t0}\\
    &t_0+\sum_{m=1}^Me^{u_m}\leq T_{\mathrm{max}},\label{P2_T}\\
    &v_m \leq \log P_{\mathrm{max}},~\forall m\in\mathcal{M},\label{P2_vm}\\
    & u_m + v_m \le \log(E_m(t_0)),~\forall m\in\mathcal{M},\label{P2_energy}\\
    & \widetilde{F}_n(\mathbf{v}) \le 0,~\forall n\in\mathcal{N},\label{P2_F_n}
    \end{align}
\end{subequations}
where $\mathbf{u} \triangleq [u_1,\ldots,u_M]^{\mathsf T}\in\mathbb{R}^{M}$ and $\mathbf{v} \triangleq [v_1,\ldots,v_M]^{\mathsf T}\in\mathbb{R}^{M}$. Compared with the original problem $(\mathcal{P}1)$, problem $(\mathcal{P}2)$ is considerably more tractable due to the concave objective function and convex constraints, except for \eqref{P2_F_n}. To tackle this issue, we construct suitable convex approximations and develop an SCA-based solution framework in the following subsection~\cite{SCA}.

\subsection{Iterative Algorithm}
\vspace{-0.1cm}
To handle the remaining nonconvexity in \eqref{P2_F_n}, we adopt an SCA-based approach by constructing a convex surrogate of $\widetilde{F}_n(\mathbf v)$. At the $r$-th iteration, let $\mathbf v^{(r)}$ denote the current feasible point. Note that the nonconvexity of $\widetilde{F}_n(\mathbf v)$ stems from the last two concave terms. Since a concave function admits a global upper bound given by its first-order Taylor expansion,
we upper-bound these concave terms at $\mathbf v^{(r)}$, which leads to an affine upper bound and thus a convex 
$\widetilde{F}_n^{(r)}(\mathbf v)$ satisfying $\widetilde{F}_n(\mathbf v)\le \widetilde{F}_n^{(r)}(\mathbf v)$ for all $\mathbf v$. The resulting surrogate is given in \eqref{F_n_r}. %% at the top of the next page.
%we construct a convex upper bound using the first-order Taylor expansion at a local feasible point. Specifically, at the $r$-th iteration, we denote the local feasible point as $\mathbf{v}^{(r)}$. Since the last two terms are concave, their first-order Taylor approximations at $\mathbf{v}^{(r)}$ provide an affine global upper bound, which is denoted by $\widetilde{F}_n^{(r)}(\mathbf{v})$ in \eqref{F_n_r} at the top of the next page. 
\begin{table*}
    \vspace{-0.3cm}
    \begin{align}
        \label{F_n_r}
        % \widetilde{F}_n(\mathbf{v})
        % \le\;&
        % \sum_{m=0}^{M} 
        % \alpha_{m,n} e^{v_m}
        % -
        % \frac{\eta}{2}
        % \sum_{i=1}^{M}
        % \sum_{j=1}^{M}
        % \beta_{i,j,n}\,
        % e^{v_i^{(r)} + v_j^{(r)}} 
        % \Big[
        % 1 + (v_i - v_i^{(r)}) + (v_j - v_j^{(r)})
        % \Big] + \mu_n
        % \triangleq
        % \widetilde{F}_n^{(r)}(\mathbf{v}),~\forall n\in\mathcal{N}.\\
        \widetilde{F}_n(\mathbf{v})
        \le&
        \sum_{m=1}^{M} \alpha_{m,n} e^{v_m}
        +\mu_n
        -\frac{\eta}{2}
        \sum_{i=1}^{M}\sum_{j=1}^{M}
        \beta_{i,j,n}
        e^{v_i^{(r)} + v_j^{(r)}}
        \Big[1+(v_i-v_i^{(r)})+(v_j-v_j^{(r)})\Big]
        \nonumber\\
        &
        -\eta p_0
        \sum_{m=1}^{M}
        \varphi_{m,n}
        e^{v_m^{(r)}}
        \Big[1+(v_m-v_m^{(r)})\Big]
        \triangleq
        \widetilde{F}_n^{(r)}(\mathbf{v}),~\forall n\in\mathcal{N}
    \end{align}
    \hrule
    \vspace{-0.3cm}
\end{table*}

Accordingly, we obtain a convex optimization problem at iteration $r$ as
%\vspace{-0.1cm}
\begin{subequations}
    \begin{align}(\mathcal{P}2^{(r)})\!:\max _{t_0,\mathbf{u},\mathbf{v}} ~~&\min_{m\in\mathcal{M}}\{\log \widetilde{R}_m(u_m,v_m)\}\nonumber\\
    \text{ s.t. }~~
    % &t_0 >0,\\
    % &t_0+\sum_{m=1}^Me^{u_m}\leq T_{\mathrm{max}},\\
    % &v_m \leq \log P_{\mathrm{max}},~\forall m\in\mathcal{M},\\
    % & u_m + v_m \le \log(E_m(t_0)),~\forall m\in\mathcal{M},\label{P2r_energy_con}\\
    & \widetilde{F}_n^{(r)}(\mathbf{v}) \le 0,~\forall n\in\mathcal{N},\label{P2r_F_n}\\
    & \eqref{P2_t0}-\eqref{P2_energy},\nonumber
    \end{align}
\end{subequations}
which can be easily solved using convex optimization tools. The inequality in \eqref{F_n_r} ensures that the constraint \eqref{P2r_F_n} is always stricter than \eqref{P2_F_n}, thus the feasible point in $(\mathcal{P}2^{(r)})$ is always feasible in $(\mathcal{P}2)$. The overall SCA-based iterative procedure %that successively solves $(\mathcal{P}2^{(r)})$ 
is summarized in Algorithm~\ref{algo1}.
\setlength{\textfloatsep}{1pt}
\begin{algorithm}[!t]%\small
\footnotesize
\algsetup{linenosize=\large}
%\vspace{.05in}
\caption{\bf{Proposed Iterative Algorithm for $(\mathcal{P} 1)$}}%\label{al:al1}
\begin{algorithmic}
\STATE \noindent{\bf{$\!\!\!\!\!$Initialization}} \\
\STATE  Initialize a feasible point $(t_0^{(0)},\mathbf{u}^{(0)},\mathbf{v}^{(0)})$, set iteration index $r=0$ and threshold $\lambda_{\mathrm{th}}$.
%\STATE $r=0.$
 \STATE \noindent{ \bf{$\!\!\!\!\!\!\!$Iteration}} \\

 \STATE \noindent{\bf{a)}} ~Build $\widetilde{F}_n^{(r)}(\mathbf v)$ and corresponding convex problem ($\mathcal{P} 2^{(r)}$) using $\mathbf v^{(r)}$;\\

 \STATE \noindent{\bf{b)}} ~Solve problem ($\mathcal{P} 2^{(r)}$) to obtain $(t_0^{(r\star)},\mathbf u^{(r\star)},\mathbf v^{(r\star)})$;\\

 \STATE \noindent{\bf{c)}} ~ {\bf{If}} the relative objective improvement is below the threshold $\lambda_{\mathrm{th}}$\\

 \STATE ~~~~~~~~~~~Define %the final solution as
 $(t_0^{\star},\mathbf{u}^{\star},\mathbf{v}^{\star})=(t_0^{(r\star)},\mathbf u^{(r\star)},\mathbf v^{(r\star)})$ and go to {\bf{d)}}.\\

 \STATE ~~~~ \noindent{\bf{Else}} 
  \STATE ~~~~~~~~~~~Define $(t_0^{(r+1)},\mathbf u^{(r+1)},\mathbf v^{(r+1)})
=(t_0^{(r\star)},\mathbf u^{(r\star)},\mathbf v^{(r\star)})$, \\
\STATE ~~~~~~~~~~~$r=r+1$, and go back to {\bf{a)}}.

 \STATE \noindent{\textbf{$\!\!\!\!\!$Reconstruction}} \\

 \STATE \noindent{\bf{d)}} ~Recover %the original variables by
$t_m^\star = e^{u_m^\star}$ and $p_m^\star = e^{v_m^\star}$, $\forall m\in\mathcal M$, and calculate the corresponding objective value.%the optimum of $(\mathcal{P} 1)$.% using the final solution $(\mathbf{u}^{\star},\mathbf{v}^{\star})$.
\end{algorithmic}
\label{algo1}
%\vspace{-.1cm}
\end{algorithm}

\section{Numerical Results}
In this section, numerical results are presented to evaluate the performance of the proposed wireless powered ISAC system. We consider an ISAC system consisting of one BS, $M=10$ users, and $N=10$ targets, where the BS is located at the origin and users as well as targets are randomly and independently distributed over a circular area of radius $10$~m centered at the BS. The wireless channel gain %between any pair of nodes 
%follows a distance-dependent path loss model 
is given by $h = z \kappa d^{-\nu}$, where $d$ denotes the distance, $\kappa = 10^{-3}$ is the reference channel gain, $\nu = 2.5$ is the path loss exponent, and $z$ represents a Rayleigh-distributed small-scale fading factor. Unless otherwise stated, the default simulation parameters are set as follows: $c = 3 \times 10^{8}$~m/s, $T_{\mathrm{max}} = 10$~s, $p_0 = 10$~W, $P_{\mathrm{max}} = 2$~W, $\sigma^2 = -70$~dBm, $\eta = 5 \times 10^{-2}$, $W = 1$~MHz, $\zeta_m = 0.7$, $\forall m \in \mathcal{M}$, and $\lambda_{\mathrm{th}}=10^{-5}$.

To further demonstrate the effectiveness of the proposed joint resource allocation scheme, we provide the following two benchmark schemes for performance comparison:
\begin{itemize}
    \item \textbf{Equal ISAC Time}: The total ISAC transmission duration is equally divided among all users, i.e., $t_m = t$, $\forall m \in \mathcal{M}$, while the variables $(t_0, t, \mathbf{p})$ are jointly optimized using a simplified variant of Algorithm~1.
    
    %The energy harvesting duration $t_0$, the common ISAC transmission time $t$, and the transmit power $\mathbf{p}$ are jointly optimized using the iterative algorithm.

    \item \textbf{Maximum Transmit Power}: All users transmit with the maximum allowable power, i.e., $p_m = P_{\max}$,~$\forall m \in \mathcal{M}$, while the variables $(t_0,\mathbf{t})$ are jointly optimized using a simplified variant of Algorithm~1.
    
    %while the energy harvesting duration $t_0$ and the ISAC transmission durations $\mathbf{t}$ are jointly optimized using a similar iterative algorithm.
\end{itemize}

\begin{figure}[!t]
	\centering
	\includegraphics[width=0.4\textwidth, trim=10 20 10 20]{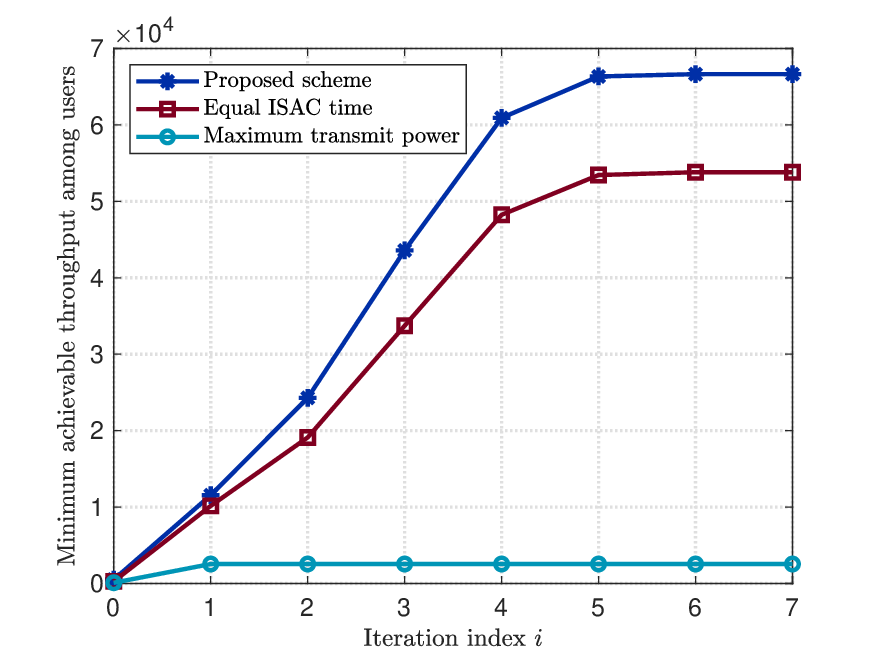}
	\caption{Convergence behavior of the proposed and benchmark schemes.}
    \label{fig_converge}
	%\vspace{-0.1cm}
\end{figure}

%The simulation results are obtained based on a random realization of the ISAC system, and similar trends are observed in other realizations. 
At first, we examine the convergence behavior of the proposed iterative algorithm. 
Fig.~\ref{fig_converge} illustrates the evolution of the minimum achievable throughput among users versus the iteration index. 
It is observed that the proposed algorithm converges rapidly and stabilizes within a few iterations, demonstrating its reliability and efficiency. 
Compared with the benchmark schemes, the proposed method achieves a higher throughput.
\begin{figure}[!t]
	\centering
	\includegraphics[width=0.4\textwidth, trim=10 20 10 20]{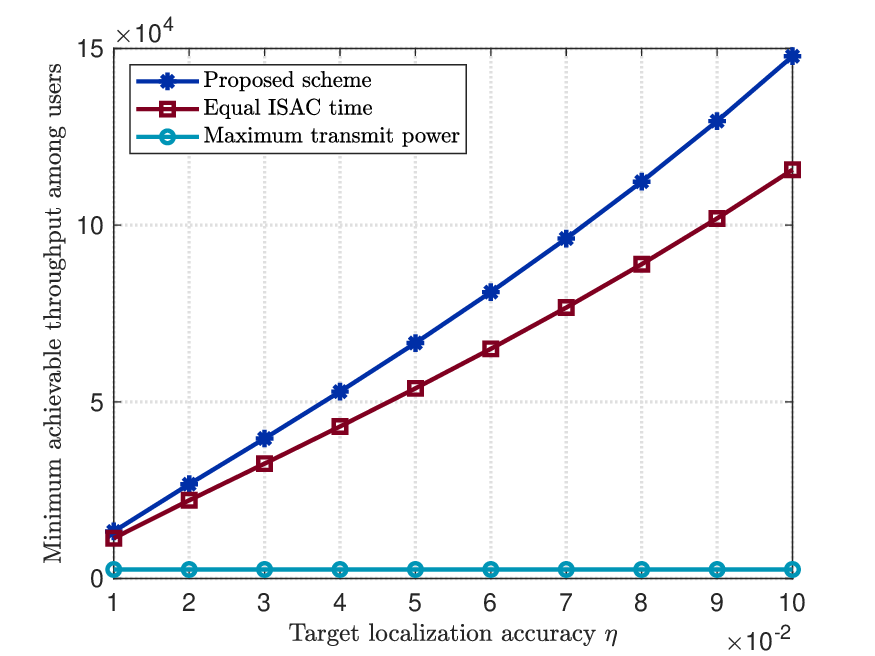}
	\caption{Minimum achievable throughput versus target localization accuracy requirement.}
    \label{fig_varying_eta}
	%\vspace{-0.6cm}
\end{figure}

Then, we evaluate how the system performance varies with the target localization accuracy requirement~$\eta$. As shown in Fig.~\ref{fig_varying_eta}, the achievable throughput of both the proposed joint optimization scheme and the equal ISAC time scheme increases with $\eta$, since a higher $\eta$ corresponds to a more relaxed localization constraint. In contrast, the maximum transmit power benchmark scheme remains insensitive to $\eta$ whenever it is feasible, because the CRB constraint depends only on the fixed transmit powers and sensing geometry. When $\eta$ is sufficiently small, the localization requirement cannot be satisfied under $P_{\max}$, and this benchmark becomes infeasible. Overall, the proposed method outperforms the benchmark schemes in varying localization requirements, highlighting the benefit of jointly optimizing the energy harvesting duration, ISAC transmission time, and transmit power.

\begin{figure}[!t]
	\centering
	\includegraphics[width=0.4\textwidth, trim=10 20 10 20]{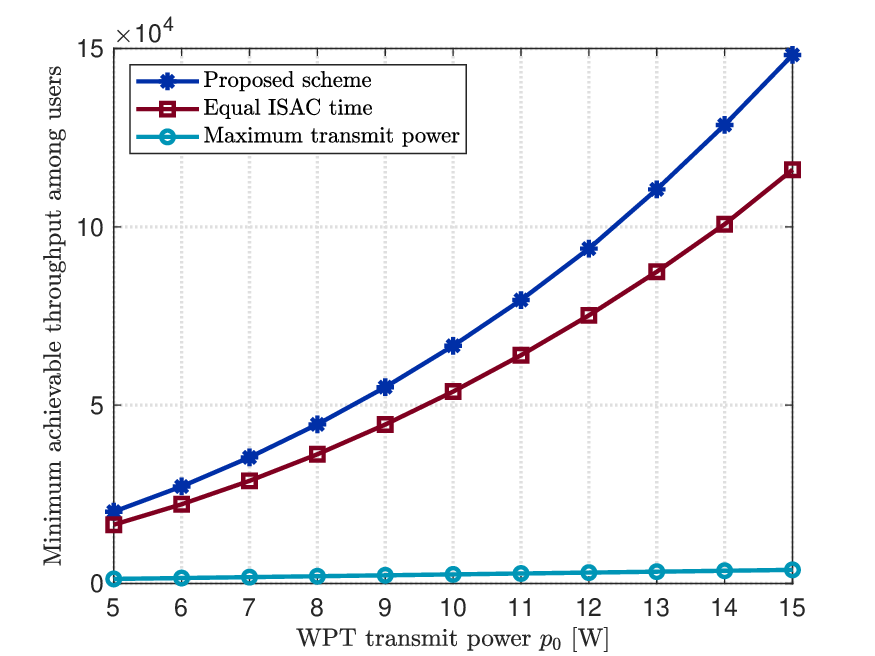}
	\caption{Minimum achievable throughput versus WPT transmit power.}
    \label{fig_varying_p0}
	%\vspace{-0.1cm}
\end{figure}

Finally, we investigate the impact of the WPT transmit power on the achievable throughput. Fig.~\ref{fig_varying_p0} shows the minimum achievable throughput among users as a function of the WPT transmit power $p_0$. As $p_0$ increases, the achievable throughput of all schemes improves, since more energy can be harvested during the WPT phase to support subsequent ISAC transmissions under the same total time budget. The proposed method consistently outperforms the benchmark schemes over the entire range of $p_0$, demonstrating the benefit of joint resource allocation. In contrast, the equal ISAC time scheme exhibits inferior performance due to its limited flexibility in time allocation. 
Moreover, the maximum transmit power scheme achieves significantly lower throughput since fixing the user transmit power reduces the degrees of freedom to balance energy harvesting, localization, and communication.

\section{Conclusion}

This paper investigates a wireless powered ISAC system with CRB-based target localization requirements. We propose a joint resource allocation framework that balances harvested energy, communication throughput, and localization accuracy. To tackle the resulting nonconvex problem, we first reformulate it via variable substitutions and logarithmic trans-
formations, then develop an efficient SCA-based iterative algorithm. Numerical results demonstrate fast convergence and %significant
consistent %performance 
throughput gains over benchmark schemes under varying localization requirements and energy-harvesting conditions, highlighting the effectiveness of the proposed joint time and power allocation in wireless powered ISAC systems. %Future work
% Our work paves the way %and is expected to be extended to %will 
% for extending to%the proposed framework to 
% more general network architectures, such as multi-BS deployments and cooperative ISAC systems.
The proposed framework paves the way for extensions to more general network architectures, such as multi-BS deployments and cooperative ISAC systems.

% Future work may extend the proposed framework to more general network scenarios, such as multi-target sensing, dynamic channel conditions, and adaptive WPT strategies.

% \appendices
% \section{Proof of Concavity of $f(v_m)$}

\bibliographystyle{IEEEtran}
%\vspace{-.1cm}
%\bibliography{IEEEabrv,D:/3.0/8.0/myref}

\begin{thebibliography}{10}
%\vspace{-.1cm}
\providecommand{\url}[1]{#1}
\csname url@samestyle\endcsname
\providecommand{\newblock}{\relax}
\providecommand{\bibinfo}[2]{#2}
\providecommand{\BIBentrySTDinterwordspacing}{\spaceskip=0pt\relax}
\providecommand{\BIBentryALTinterwordstretchfactor}{4}
\providecommand{\BIBentryALTinterwordspacing}{\spaceskip=\fontdimen2\font plus
\BIBentryALTinterwordstretchfactor\fontdimen3\font minus
  \fontdimen4\font\relax}
\providecommand{\BIBforeignlanguage}[2]{{%
\expandafter\ifx\csname l@#1\endcsname\relax
\typeout{** WARNING: IEEEtran.bst: No hyphenation pattern has been}%
\typeout{** loaded for the language `#1'. Using the pattern for}%
\typeout{** the default language instead.}%
\else
\language=\csname l@#1\endcsname
\fi
#2}}
\providecommand{\BIBdecl}{\relax}
\BIBdecl
%\vspace{-.1cm}
% \bibitem{Trans_1}


%% Introduction


\bibitem{ISAC_advance}
S. Lu et al., ``Integrated Sensing and Communications: Recent Advances and Ten Open Challenges,'' \emph{IEEE Internet Things J.}, vol. 11, no. 11, pp. 19094-19120, Jun. 2024.

\bibitem{ISAC_application}
Y. Cui, F. Liu, X. Jing and J. Mu, ``Integrating Sensing and Communications for Ubiquitous IoT: Applications, Trends, and Challenges,'' \emph{IEEE
Netw.}, vol. 35, no. 5, pp. 158-167, Sep. 2021.

% \bibitem{ISAC_waveform}
% Z. Xiao and Y. Zeng, ``Waveform Design and Performance Analysis for Full-Duplex Integrated Sensing and Communication,'' \emph{IEEE J. Sel. Areas
% Commun.}, vol. 40, no. 6, pp. 1823-1837, Jun. 2022.
\bibitem{ISAC_waveform1}
S. Wang, W. Dai, H. Wang and G. Y. Li, ``Robust Waveform Design for Integrated Sensing and Communication,'' \emph{IEEE
Trans. Signal Process.}, vol. 72, pp. 3122-3138, Jun. 2024.

\bibitem{ISAC_waveform2}
P. Wang, D. Han, Y. Cao, W. Ni and D. Niyato, ``Multi-Objective Optimization-Based Waveform Design for Multi-User and Multi-Target MIMO-ISAC Systems,'' \emph{IEEE Trans. Wireless Commun.}, vol. 23, no. 10, pp. 15339-15352, Oct. 2024.

\bibitem{ISAC_multiple_access}
L. Sun, Z. Zhao, S. Wang, Z. Ding and M. Peng, ``On the Study of Non-Orthogonal Multiple Access (NOMA)-Assisted Integrated Sensing and Communication (ISAC),'' \emph{IEEE Trans. Commun.}, vol. 72, no. 11, pp. 7278-7293, Nov. 2024.

\bibitem{ISAC_dual}
F. Liu et al., ``Integrated Sensing and Communications: Toward Dual-Functional Wireless Networks for 6G and Beyond,'' \emph{IEEE J. Sel. Areas
Commun.}, vol. 40, no. 6, pp. 1728-1767, Jun. 2022.

\bibitem{ISAC_localization1}
Z. Zhang et al., ``Target Localization in Cooperative ISAC Systems: A Scheme Based on 5G NR OFDM Signals,'' \emph{IEEE Trans. Commun.}, vol. 73, no. 5, pp. 3562-3578, May 2025.

\bibitem{ISAC_localization2}
X. Jing, F. Liu, C. Masouros and Y. Zeng, ``ISAC From the Sky: UAV Trajectory Design for Joint Communication and Target Localization,'' \emph{IEEE Trans. Wireless Commun.}, vol. 23, no. 10, pp. 12857-12872, Oct. 2024.




% \bibitem{ISAC_signal}
% J. A. Zhang et al., ``An Overview of Signal Processing Techniques for Joint Communication and Radar Sensing,'' \emph{IEEE J. Sel. Topics Signal Process.}, vol. 15, no. 6, pp. 1295-1315, Nov. 2021.


% \bibitem{ISAC_OTFS}
% N. Wu, H. Li, D. He, A. Nallanathan and T. Q. S. Quek, ``Integrated Sensing and Communication Receiver Design for OTFS-Based MIMO System: A Unified Variational Inference Framework," \emph{IEEE J. Sel. Areas
% Commun.}, vol. 43, no. 4, pp. 1339-1353, Apr. 2025.








% \bibitem{ISCP_magazine}
% X. Li et al., ``Integrating Sensing, Communication, and Power Transfer: From Theory to Practice,'' \emph{IEEE Communications Magazine}, vol. 62, no. 9, pp. 122-127, Sept. 2024.
% \bibitem{estimation_book}
% S. M. Kay, \emph{Fundamentals of Statistical Signal Processing: Estimation Theory}, Englewood Cliffs, NJ, USA: Prentice-Hall, 1993.

% \bibitem{MSE_CRB}
% A. Liu et al., ``A Survey on Fundamental Limits of Integrated Sensing and Communication,'' \emph{IEEE Commun. Surveys Tuts.}, vol. 24, no. 2, pp. 994-1034, 2nd Quart., 2022.



\bibitem{ISCPT}
X. Li et al., ``Integrating Sensing, Communication, and Power Transfer: From Theory to Practice,'' \emph{IEEE Commun. Mag.}, vol. 62, no. 9, pp. 122-127, Sep. 2024.

\bibitem{ISCAP}
Y. Chen, Z. Ren, J. Xu, Y. Zeng, D. W. K. Ng and S. Cui, ``Integrated Sensing, Communication, and Powering: Toward Multi-Functional 6G Wireless Networks,'' \emph{IEEE Commun. Mag.}, vol. 63, no. 8, pp. 146-153, Aug. 2025.



\bibitem{wireless_ISAC}
X. Li, Z. Han, Z. Zhou, Q. Zhang, K. Huang, and Y. Gong,
“Wirelessly Powered Integrated Sensing and Communication,” in \emph{Proc. 1st ACM MobiCom Workshop ISCS}, Sydney, Australia, 2022, pp. 1-6.

\bibitem{UAV_ISAC}
O. Rezaei, M. M. Naghsh, S. M. Karbasi and M. M. Nayebi, ``Resource Allocation for UAV-Enabled Integrated Sensing and Communication (ISAC) via Multi-Objective Optimization,''
in \emph{Proc. IEEE Int. Conf. Acoust., Speech Signal Process. (ICASSP)}, Rhodes Island, Greece, 2023, pp. 1-5.




\bibitem{target_localization_con}
H. Godrich, A. Petropulu and H. V. Poor, ``Power Allocation Schemes for Target Localization in Widely Distributed MIMO Radar Systems,'' in \emph{Proc. IEEE Mil. Commun. Conf. (MILCOM)}, San Jose, CA, USA, 2010, pp. 846-851.

\bibitem{target_localization}
H. Godrich, A. P. Petropulu and H. V. Poor, ``Power Allocation Strategies for Target Localization in Distributed Multiple-Radar Architectures,'' \emph{IEEE
Trans. Signal Process.}, vol. 59, no. 7, pp. 3226-3240, Jul. 2011.

\bibitem{loglog1}
X. Yuan, Y. Hu, M. Liu, T. Matsumura and A. Schmeink, ``Optimal Beam Deployment for FSO Link Assisted Satellite-Ground Multicasting Communication,'' in \emph{Proc. IEEE Wireless Commun. and Networking Conf. (WCNC)}, Milan, Italy, 2025, pp. 1-6.

\bibitem{loglog2}
P. Zheng, B. Li, X. Yuan, Y. Hu and A. Schmeink, ``Multi-UAV-Enabled Cognitive Radio Networks: Joint UAV Deployment and Resource Allocation Design,'' in \emph{Proc. 28th Int. Workshop Smart Antennas (WSA)}, Erlangen, Germany, 2025, pp. 208-213.





\bibitem{SCA}
G. Scutari and Y. Sun, ``Parallel and Distributed Successive Convex Approximation Methods for Big-Data Optimization,'' in \emph{Multi-Agent Optimization}. Cham, Switzerland: Springer, Jan. 2018, pp. 141-308.


%% Problem Formulation










\end{thebibliography}

\bibliographystyle{IEEEtran}

%%%%%%%%%%%%%%%%%%%%%%%%%%%

\end{document}